\documentclass[conference]{IEEEtran}
\IEEEoverridecommandlockouts
\usepackage{cite}
\usepackage{amsmath,amssymb,amsfonts}
\usepackage{algorithmic}
\usepackage{graphicx}
\usepackage{textcomp}
\usepackage{xcolor}
\usepackage{multirow}
\usepackage[caption=false,font=footnotesize]{subfig}
\def\BibTeX{{\rm B\kern-.05em{\sc i\kern-.025em b}\kern-.08em
    T\kern-.1667em\lower.7ex\hbox{E}\kern-.125emX}}
\usepackage{braket}
\usepackage{tikz}
\usetikzlibrary{arrows.meta,positioning,fit,backgrounds,calc}

\begin{document}

\title{Quantum Transformer BSDE Solver via Multi-Layer Fully-Connected Variational Quantum Circuits}

\author{

\IEEEauthorblockN{Howard Su}
\IEEEauthorblockA{Imperial College London\\
London, UK \\
h.su24@imperial.ac.uk}
\and
\IEEEauthorblockN{Huan-Hsin Tseng}
\IEEEauthorblockA{Brookhaven National Laboratory
\\
NY, USA \\
htseng@bnl.gov}
\and
\IEEEauthorblockN{Chi-Sheng Chen}
\IEEEauthorblockA{Harvard Medical School\\
Boston, MA, USA \\
m50816m50816@gmail.com}
\and
\IEEEauthorblockN{Lance Bai}
\IEEEauthorblockA{Independent Researcher\\
London, UK \\
lancebai@gmail.com}
}

\maketitle

\begin{abstract}
Solving high-dimensional parabolic partial differential equations
(PDEs) is important in engineering, physics, and stochastic control.
Deep BSDE methods reformulate semilinear PDEs as backward stochastic
differential equations and admit a model-based reinforcement learning
interpretation, where trajectories are generated from known stochastic
dynamics while a trainable model learns the gradient-related control
process. We propose a Quantum Transformer BSDE solver based on
Multi-Layer Fully-Connected Variational Quantum Circuits (FC-VQC).
The method treats the normalized state trajectory as time--coordinate
tokens and applies causal self-attention to learn interactions in the
adapted BSDE gradient process. All trainable model parameters are
contained within the FC-VQC embedding, projection, feed-forward, and decoder
modules, while attention and structural operations remain classical
and parameter-free. Experiments on three \(d=36\) PDE benchmarks show
that QTransformer consistently improves over the non-attentive FC-VQC
baseline and outperforms the classical Transformer at compact hidden
widths, while the wider classical Transformer achieves the best
overall accuracy. These results demonstrate that combining causal
attention with FC-VQC provides an effective quantum architecture for
high-dimensional BSDE trajectory learning.
\end{abstract}


\section{Introduction}
Solving high-dimensional parabolic partial differential equations
(PDEs) is important in engineering, physics, and stochastic
control. Classical grid-based methods suffer from the curse of
dimensionality, while Deep BSDE methods
\cite{e2017deep}\cite{han2018solving} reformulate semilinear PDEs as
backward stochastic differential equations (BSDE)
\cite{pardoux1992backward}\cite{pardoux1999forward}. This gives a
model-based reinforcement learning viewpoint: trajectories are
generated from known stochastic dynamics, and the trainable model
learns the gradient-related control process. The main
learning task is therefore to approximate the map
\(X_{t_n}\mapsto Z_{t_n}\), which drives the backward propagation of
the PDE solution.

Variational quantum circuits (VQCs) are widely used trainable
models in quantum machine learning, where classical data are encoded
into quantum states, processed by parameterized gates, and read out
through measurement expectations
\cite{benedetti2019parameterized}\cite{cerezo2021variational}. However,
VQC design faces an expressivity--trainability trade-off: shallow
circuits may lack capacity, while wider or deeper circuits increase
simulation cost and barren-plateau risk
\cite{mcclean2018barren}\cite{cerezo2021cost}. A previous quantum BSDE
solver used VQC-based maps to approximate the BSDE gradient process
\(Z_{t_n}\) without trainable classical neural-network layers
\cite{su2025quantumbsde}. FC-VQC extends this idea by partitioning
high-dimensional inputs into local VQC blocks and propagating
information through deterministic shifted-ring mixing
\cite{su2026fcvqc}. This modularity is especially useful for
Transformer-style models: larger hidden dimensions
\(d_{\mathrm{model}}\) can be obtained by adding local VQC blocks,
rather than increasing the qubit count of a monolithic VQC.

Recent quantum Transformer studies\cite{zhang2025surveyQuantumTransformers} include near-term PQC-based hybrid
models\cite{li2024quantumSelfAttention}, where quantum circuits generate Transformer representations
while attention remains classical.
Related work on Quantum Adaptive Self-Attention further highlights the
importance of capacity-matched classical controls when assessing
quantum Transformer gains \cite{chen2025qasa}. We introduce a Quantum
Transformer BSDE solver based on FC-VQC. The key idea is to replace every trainable map in a Transformer
\cite{vaswani2017attention} with local FC-VQC modules. Thus, all
trainable model parameters are contained within the FC-VQC modules,
while attention, softmax, causal masking, residual connections,
concatenation, and reshaping remain classical and parameter-free. Our contributions are: (i) causal self-attention for time--space
BSDE trajectory modelling; (ii) a QTransformer whose trainable model
parameters are confined to scalable FC-VQC modules; and (iii) an
evaluation against classical and quantum baselines at multiple hidden
widths on Black--Scholes, Burgers-type, and oscillatory
reaction--diffusion PDEs.

%\enlargethispage{2\baselineskip}
\section{QUANTUM TRANSFORMER BSDE SOLVER}

We present the proposed solver in three steps: the BSDE representation
of semilinear parabolic PDEs, the FC-VQC building block, and the
FC-VQC-based Quantum Transformer architecture for learning \(Z_{t_n}\).

\subsection{BSDE Reformulation}
We consider a semilinear parabolic PDE on
\([0,T]\times\mathbb{R}^d\) for the unknown solution
\(u:[0,T]\times\mathbb{R}^d\to\mathbb{R}\):
\begin{equation}\label{eq:pde}
\begin{aligned}
&\partial_t u(t,x)+\mathcal{L}u(t,x)
+f\!\left(t,x,u(t,x),\sigma(t,x)^\top\nabla_xu(t,x)\right)=0,\\
&u(T,x)=g(x),
\end{aligned}
\end{equation}
where
\(\mathcal{L}u=\mu(t,x)^\top\nabla_xu+
\frac{1}{2}\mathrm{Tr}(\sigma(t,x)\sigma(t,x)^\top\nabla_x^2u)\)
is the infinitesimal generator, \(T\) is the terminal time,
\(d\) is the spatial dimension,
\(\mu(t,x)\in\mathbb{R}^d\) is the drift,
\(\sigma(t,x)\in\mathbb{R}^{d\times d}\) is the diffusion matrix, and
\(f:[0,T]\times\mathbb{R}^d\times\mathbb{R}\times\mathbb{R}^d
\to\mathbb{R}\) is the driver.

By the nonlinear Feynman--Kac formula, \eqref{eq:pde} admits the
following BSDE representation:
\begin{equation}
\begin{aligned}
\mathrm{d}X_t &= \mu(t,X_t)\mathrm{d}t
+\sigma(t,X_t)\mathrm{d}W_t, \qquad X_0=\xi,\\
\mathrm{d}Y_t &= -f(t,X_t,Y_t,Z_t)\mathrm{d}t
+Z_t^\top\mathrm{d}W_t, \qquad Y_T=g(X_T),
\end{aligned}
\label{eq:fbsde}
\end{equation}
where \(Y_t=u(t,X_t)\) and
\(Z_t=\sigma(t,X_t)^\top\nabla_xu(t,X_t)\).

This gives a model-based reinforcement learning viewpoint: the forward
SDE is the known transition model, and \(\widehat Z_{t_n}\) acts as a
policy-like control in the backward value recursion.

After discretizing \(0=t_0<\cdots<t_N=T\), the solver simulates
\(X_{t_n}\) by Euler--Maruyama and propagates
\begin{equation}
Y_{t_{n+1}}
=
Y_{t_n}
-
f(t_n,X_{t_n},Y_{t_n},\widehat Z_{t_n})\Delta t
+
\widehat Z_{t_n}^{\top}\Delta W_n .
\label{eq:discrete_bsde}
\end{equation}
Quantum Transformer predicts the full gradient trajectory
\begin{equation}
(\widehat Z_{t_0},\ldots,\widehat Z_{t_{N-1}})
=
\mathcal{T}_{\Theta}
(\bar X_{t_0},\ldots,\bar X_{t_{N-1}}),
\label{eq:z_prediction}
\end{equation}
where \(\bar X_{t_n}\) denotes the normalized model input. The trainable
parameters \(\Theta\), together with the initial scalar \(Y_0=y_0\), are
optimized by the terminal loss
\begin{equation}
\mathcal{L}(\Theta,y_0)
=
\mathbb{E}\left[
\left|Y_{t_N}-g(X_{t_N})\right|^2
\right].
\label{eq:terminal_loss}
\end{equation}

\subsection{FC-VQC Architecture}

We use FC-VQC to replace dense trainable maps. Instead of encoding a \(d\)-dimensional
vector into one \(d\)-qubit circuit, FC-VQC partitions the input into
fixed-size local VQC blocks. For \(x\in\mathbb{R}^d\), we zero-pad the
input, if necessary, to \(\tilde d=Bq\), where \(q\) is the number of
qubits per local block and \(B\) is the number of blocks:
\begin{equation}
\tilde{x}=[x^{(1)},\ldots,x^{(B)}],
\qquad x^{(b)}\in\mathbb{R}^{q}.
\label{eq:fcvqc_partition}
\end{equation}

Each block is processed by an independent local VQC
\(f_{\phi_b^{(\ell)}}:\mathbb{R}^{q}\rightarrow\mathbb{R}^{q}\),
consisting of rotation encoding, \(K\) parameterized entangling layers,
and Pauli-\(Z\) expectation measurements. At FC-VQC layer \(\ell\),
parameter-free shifted-ring mixing is first applied across neighbouring
blocks, and the mixed block is then passed through a local VQC:
\begin{equation}
\begin{aligned}
\tilde h^{(\ell,b)}
&=g_b^{(\ell)}
\left(h^{(\ell,1)},\ldots,h^{(\ell,B)}\right),\\
h^{(\ell+1,b)}
&=f_{\phi_b^{(\ell)}}\left(\tilde h^{(\ell,b)}\right),
\qquad b=1,\ldots,B .
\end{aligned}
\label{eq:fcvqc_layer}
\end{equation}
For the \(q=3\) blocks used in our BSDE/PDE experiments, the shifted-ring
mixer can be written, with cyclic block indices as
\[
\tilde h^{(\ell,b)}
=
\left(
h^{(\ell,b-1)}_1,\,
h^{(\ell,b)}_2,\,
h^{(\ell,b+1)}_3
\right).
\]

Increasing the width therefore adds local VQC blocks rather than
qubits per circuit, avoiding a monolithic
\(d_{\mathrm{model}}\)-qubit VQC whose simulation cost grows
exponentially with \(d_{\mathrm{model}}\).

Stacking these layers gives a nonlinear dimension-preserving map
whose trainable parameters scale linearly with the input dimension. We use an FC-VQC reducer with
one-qubit readout: each local \(q=3\) VQC block outputs one scalar,
and the resulting features are further reduced to the target dimension.
See \cite{su2026fcvqc} for the full FC-VQC construction.

\subsection{FC-VQC-Based Quantum Transformer}
% Add to preamble:
% \usepackage{tikz}
% \usetikzlibrary{arrows.meta,positioning,fit,backgrounds}
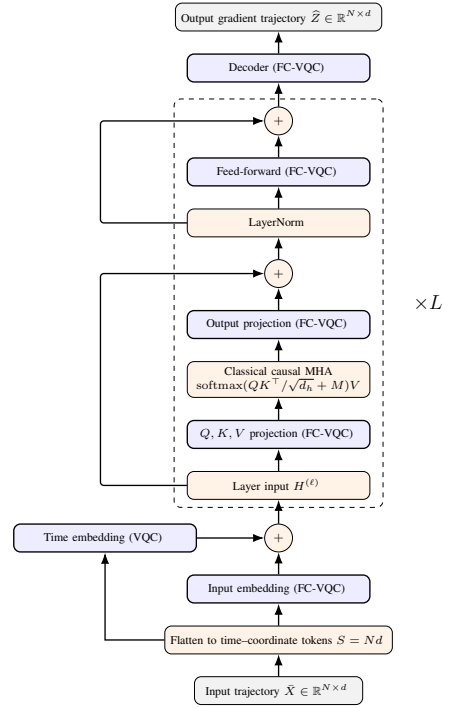
\begin{figure}[t]
\centering
\resizebox{0.65\columnwidth}{!}{%
\begin{tikzpicture}[
    font=\scriptsize,
    >=Latex,
    node distance=4.6mm,
    data/.style={
        draw,
        rounded corners,
        align=center,
        fill=gray!10,
        minimum width=35mm,
        minimum height=5.8mm
    },
    qblock/.style={
        draw,
        rounded corners,
        align=center,
        fill=blue!7,
        minimum width=38mm,
        minimum height=5.8mm,
        thick
    },
    cblock/.style={
        draw,
        rounded corners,
        align=center,
        fill=orange!10,
        minimum width=38mm,
        minimum height=5.8mm
    },
    add/.style={
        draw,
        circle,
        align=center,
        fill=orange!10,
        minimum size=5.8mm,
        inner sep=0pt
    },
    group/.style={
        draw,
        rounded corners,
        dashed,
        inner xsep=2.6mm,
        inner ysep=1.6mm
    },
    arrow/.style={->, thick},
    resarrow/.style={->, thick, rounded corners=2pt}
]

% ------------------------------------------------------------------
% Main route
% ------------------------------------------------------------------

\node[data] (input) {
Input trajectory \(\bar X\in\mathbb{R}^{N\times d}\)
};

\node[cblock, above=of input] (tokens) {
Flatten to time--coordinate tokens \(S=Nd\)
};

\node[qblock, above=of tokens] (embed) {
Input embedding (FC-VQC)
};

\node[add, above=of embed] (addtime) {\(+\)};

% Classical layer input inside the repeated Transformer block
\node[cblock, above=5mm of addtime] (hstate) {
Layer input \(H^{(\ell)}\)
};

\node[qblock, above=of hstate] (qkv) {
\(Q,K,V\) projection (FC-VQC)
};

\node[cblock, above=of qkv] (mha) {
Classical causal MHA\\
\(\mathrm{softmax}(QK^\top/\sqrt{d_h}+M)V\)
};

\node[qblock, above=of mha] (oproj) {
Output projection (FC-VQC)
};

\node[add, above=of oproj] (addattn) {\(+\)};

\node[cblock, above=of addattn] (layerNorm) {
LayerNorm
};

\node[qblock, above=of layerNorm] (ffn) {
Feed-forward (FC-VQC)
};

\node[add, above=of ffn] (addffn) {\(+\)};

% Add extra space after the repeated Transformer block
\node[qblock, above=5mm of addffn] (decoder) {
Decoder (FC-VQC)
};

\node[data, above=of decoder] (output) {
Output gradient trajectory \(\widehat Z\in\mathbb{R}^{N\times d}\)
};

% ------------------------------------------------------------------
% Time embedding side branch
% ------------------------------------------------------------------

\node[qblock, left=14mm of addtime] (timeemb) {
Time embedding (VQC)
};

% ------------------------------------------------------------------
% Main arrows
% ------------------------------------------------------------------

\draw[arrow] (input) -- (tokens);
\draw[arrow] (tokens) -- (embed);
\draw[arrow] (embed) -- (addtime);

% Time embedding side branch
\draw[arrow] (tokens.west) -| (timeemb.south);
\draw[arrow] (timeemb.east) -- (addtime.west);

\draw[arrow] (addtime) -- (hstate);
\draw[arrow] (hstate) -- (qkv);
\draw[arrow] (qkv) -- (mha);
\draw[arrow] (mha) -- (oproj);
\draw[arrow] (oproj) -- (addattn);
\draw[arrow] (addattn) -- (layerNorm);
\draw[arrow] (layerNorm) -- (ffn);
\draw[arrow] (ffn) -- (addffn);
\draw[arrow] (addffn) -- (decoder);
\draw[arrow] (decoder) -- (output);

% ------------------------------------------------------------------
% Classical residual connections outside the main route
% ------------------------------------------------------------------

% Residual around causal MHA + output projection:
% H^{(\ell)} + O(MHA(H^{(\ell)}))
\draw[resarrow]
    (hstate.west) -- ++(-18mm,0) |- (addattn.west);

% Residual around feed-forward block:
% \bar H^{(\ell)} + FF(\bar H^{(\ell)})
\draw[resarrow]
    (layerNorm.west) -- ++(-18mm,0) |- (addffn.west);

% ------------------------------------------------------------------
% Repeated Transformer block
% ------------------------------------------------------------------

\begin{scope}[on background layer]
\node[group,
      fit=(hstate)(qkv)(mha)(oproj)(addattn)(layerNorm)(ffn)(addffn)]
      (trblock) {};
\end{scope}

\node[anchor=west] at ([xshift=5mm]trblock.east) {\large{\textbf{\(\times L\)}}};

\end{tikzpicture}%
}
\caption{The Quantum Transformer model architecture.}
\vspace{-1em}
\label{fig:qtransformer_flowchart}
\end{figure}

For one simulated path, let $\bar X=(\bar X_{t_0},\ldots,\bar X_{t_{N-1}})
\in\mathbb{R}^{N\times d}$
be the normalized state trajectory. The model learns the
pathwise map $\widehat Z
=
\mathcal{T}_{\Theta}(\bar X),
\mathcal{T}_{\Theta}:\mathbb{R}^{N\times d}
\rightarrow
\mathbb{R}^{N\times d},
\label{eq:qtrans_map}
$
where \(\widehat Z\) is used in the discrete BSDE recursion.

Each scalar component \(\bar X_{t_n}^{(i)}\) is treated as a
time--coordinate token. It is repeated to \(d_{\mathrm{model}}\)
dimensions and passed through a width-preserving FC-VQC embedding:
\begin{equation}
H_{n,i}^{(0)}
=
\mathcal{F}_{\theta_{\mathrm{emb}}}
\left(
\bar X_{t_n}^{(i)}
\mathbf{1}_{d_{\mathrm{model}}}
\right)
+
\eta\, e_n^{\mathrm{time}},
\label{eq:qtrans_embedding}
\end{equation}
where \(e_n^{\mathrm{time}}\) is an optional quantum-generated time
embedding and \(\eta\in\{0,1\}\) indicates whether it is used. No
coordinate or asset embedding is used in the BSDE solver. The
\(N\times d\) tokens are then flattened into a sequence of length
\(S=Nd\).

For Transformer layer \(\ell=0,\ldots,L-1\), FC-VQC modules generate the query, key,
and value representations:
\begin{equation}
Q^{(\ell)}
=
\mathcal{F}_{\theta_Q^{(\ell)}}(H^{(\ell)}),\;
K^{(\ell)}
=
\mathcal{F}_{\theta_K^{(\ell)}}(H^{(\ell)}),\;
V^{(\ell)}
=
\mathcal{F}_{\theta_V^{(\ell)}}(H^{(\ell)}).
\label{eq:qkv_fcvqc}
\end{equation}

Consequently, a larger Transformer hidden size is realised by more
small FC-VQC blocks, whereas a monolithic quantum Transformer would
require larger \(d_{\mathrm{model}}\)-qubit circuits in each trainable
map.

The representations are split into \(n_h\) heads with
\(d_h=d_{\mathrm{model}}/n_h\). For each head, we apply scaled
dot-product attention:
\begin{equation}
A_h^{(\ell)}
=
\mathrm{softmax}
\left(
\frac{Q_h^{(\ell)}K_h^{(\ell)\top}}{\sqrt{d_h}}
+
M
\right),
\qquad
U_h^{(\ell)}
=
A_h^{(\ell)}V_h^{(\ell)} .
\label{eq:qtrans_attention}
\end{equation}
The mask \(M\) is causal in time, so each token can attend only to
tokens from the current and previous time steps, preserving the adapted
structure of the BSDE.

The head outputs are concatenated, $U^{(\ell)}=\mathrm{Concat}_{h=1}^{n_h} U_h^{(\ell)}$, projected by an FC-VQC module $\mathcal{F}_{\theta_O^{(\ell)}}$, and
combined with a single FC-VQC residual feed-forward $\mathcal{F}_{\theta_{\mathrm{ff}}^{(\ell)}}$:
\begin{equation}
\begin{aligned}
H^{(\ell+1)}
=
H^{(\ell)}
+
\mathcal{F}_{\theta_O^{(\ell)}}
\left(
U^{(\ell)}
\right)
+
\mathcal{F}_{\theta_{\mathrm{ff}}^{(\ell)}}
\left(
\bar H^{(\ell)}
\right).
\end{aligned}
\label{eq:qtrans_output_ffn}
\end{equation}

After the final Transformer layer, a VQC-based decoder maps each token
representation back to one scalar:
\begin{equation}
\widehat Z_{t_n}^{(i)}
=
\mathcal{D}_{\theta_{\mathrm{dec}}}
\left(
H_{n,i}^{(L)}
\right),
\qquad
n=0,\ldots,N-1,\quad i=1,\ldots,d .
\label{eq:qtrans_decoder}
\end{equation}
In the implementation, \(\mathcal{D}_{\theta_{\mathrm{dec}}}\) is
realized by a width-preserving FC-VQC module followed by an FC-VQC
dimension-reduction step with one-qubit local readouts.

Thus, all trainable parameters are contained within the FC-VQC
embedding, QKV and output-projection, feed-forward, and decoder
modules, while the attention and structural operations remain
classical and parameter-free.

\section{Experiments}
\label{sec:experiments}

\subsection{Benchmarks and Protocol}

We evaluate on three \(d=36\) parabolic PDE benchmarks, written as
task-specific instances of \eqref{eq:pde}--\eqref{eq:fbsde}. Let
\(D_x=\mathrm{diag}(x_1,\ldots,x_d)\) and let
\(\mathbf{1}\in\mathbb{R}^d\) denote the all-ones vector.

\subsubsection{Black--Scholes PDE/BSDE}
\begin{equation}
\begin{aligned}
&\partial_tu
+r x^\top\nabla_xu
+\frac{1}{2}\sigma_{\mathrm{BS}}^2
\mathrm{Tr}\!\left(D_x^2\nabla_x^2u\right)
-ru=0,
\\
&u(T,x)=g_{\mathrm{port}}(x),\\
&\mathrm{d}X_t
=
rX_t\,\mathrm{d}t
+
\sigma_{\mathrm{BS}}X_t\mathrm{d}W_t,
\\
&\mathrm{d}Y_t
=
rY_t\,\mathrm{d}t+Z_t^\top\mathrm{d}W_t,
\qquad
Y_T=g_{\mathrm{port}}(X_T).
\end{aligned}
\label{eq:bs_pde_bsde}
\end{equation}
Here \(\sigma_{\mathrm{BS}}\) is the constant volatility shared by all
assets, and \(g_{\mathrm{port}}\) denotes the terminal portfolio payoff.

\subsubsection{Burgers-type PDE/BSDE}
\begin{equation}
\begin{aligned}
&\partial_tu
+
\frac{d^2}{2}\mathrm{Tr}\!\left(\nabla_x^2u\right)
+
d(u-c_d)\mathbf{1}^\top\nabla_xu=0,
\\
&u(T,x)=\psi(T,x),\\
&\mathrm{d}X_t
=
d\,\mathrm{d}W_t,
\qquad
\mathrm{d}Y_t
=
-(Y_t-c_d)\mathbf{1}^{\top}Z_t\,\mathrm{d}t
+
Z_t^\top\mathrm{d}W_t,
\\
&Y_T=\psi(T,X_T),
\end{aligned}
\label{eq:burgers_pde_bsde}
\end{equation}
where $c_d=\frac{d+2}{2d}$, $\psi(t,x)=
\frac{\exp(t+d^{-1}\mathbf{1}^\top x)}
{1+\exp(t+d^{-1}\mathbf{1}^\top x)}$.
\vspace{1em}

\subsubsection{Oscillatory reaction--diffusion PDE/BSDE}
\begin{equation}
\begin{aligned}
&\partial_tu
+
\frac{1}{2}\mathrm{Tr}\!\left(\nabla_x^2u\right)
+
m(t,x,u)=0,
\\
&u(T,x)=1+\kappa+\sin(\lambda\mathbf{1}^\top x),\\
&\mathrm{d}X_t
=
\mathrm{d}W_t,
\qquad
\mathrm{d}Y_t
=
-m(t,X_t,Y_t)\mathrm{d}t
+
Z_t^\top\mathrm{d}W_t,\\
&Y_T
=
1+\kappa+\sin(\lambda\mathbf{1}^\top X_T),
\end{aligned}
\label{eq:osc_pde_bsde}
\end{equation}
where
$m(t,x,y)=\min\{1,[y-\kappa-1-a(t,x)]^2\}$,\\ $a(t,x)=\sin(\lambda\mathbf{1}^\top x)
\exp\!\left(\frac{\lambda^2d(t-T)}{2}\right)$, $\kappa=0.6$, $\lambda=d^{-1/2}$.

For evaluation, the analytical references are $u_{\mathrm{BS}}(t,x)$ is the closed-form Black--Scholes formula, $u_{\mathrm{B}}(t,x)=\psi(t,x)$ and $u_{\mathrm{O}}(t,x)=1+\kappa+a(t,x)$. 

All models use the same Euler--Maruyama paths, terminal condition, and
terminal BSDE loss, with \(T=1\), \(N=10\), 1,000 Monte Carlo paths,
batch size 128, and 10,000 epochs. We train with Adam using learning
rate \(5\times10^{-3}\) and report averages over seeds
\(\{42,123,456\}\). We sweep \(L\in\{3,5,7\}\), where \(L\) denotes
model layers for DNN/FC-VQC and Transformer blocks for
Transformer/QTransformer. For FC-VQC, the local VQC depth is set to
\(K=L\); for QTransformer, all internal FC-VQC modules use
\(L_{\mathrm{VQC}}=3\) and \(K=3\). Accuracy is measured by
$\mathrm{RelMAE}
=
\frac{1}{N+1}
\sum_{n=0}^{N}
\frac{
\mathbb{E}|Y_{t_n}-u(t_n,X_{t_n})|
}{
\mathbb{E}|u(t_n,X_{t_n})|+\varepsilon
}$.

\subsection{Results}
\label{subsec:results}
\begin{figure*}[t]
\centering
\begin{minipage}[t]{0.31\textwidth}
\centering
\includegraphics[width=\linewidth]{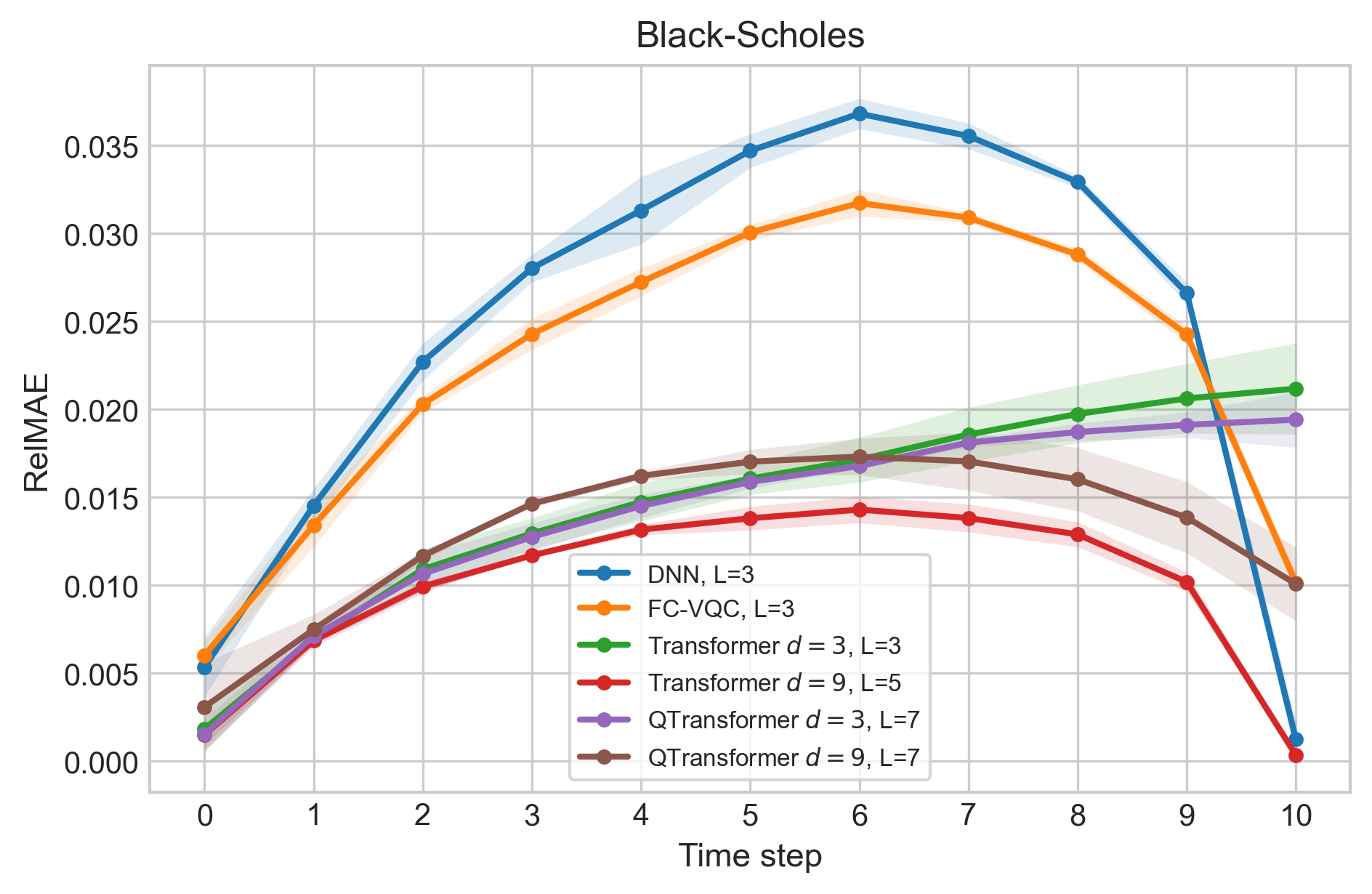}\\[-0.3em]
{\footnotesize (a) Black--Scholes}
\end{minipage}
\hfill
\begin{minipage}[t]{0.31\textwidth}
\centering
\includegraphics[width=\linewidth]{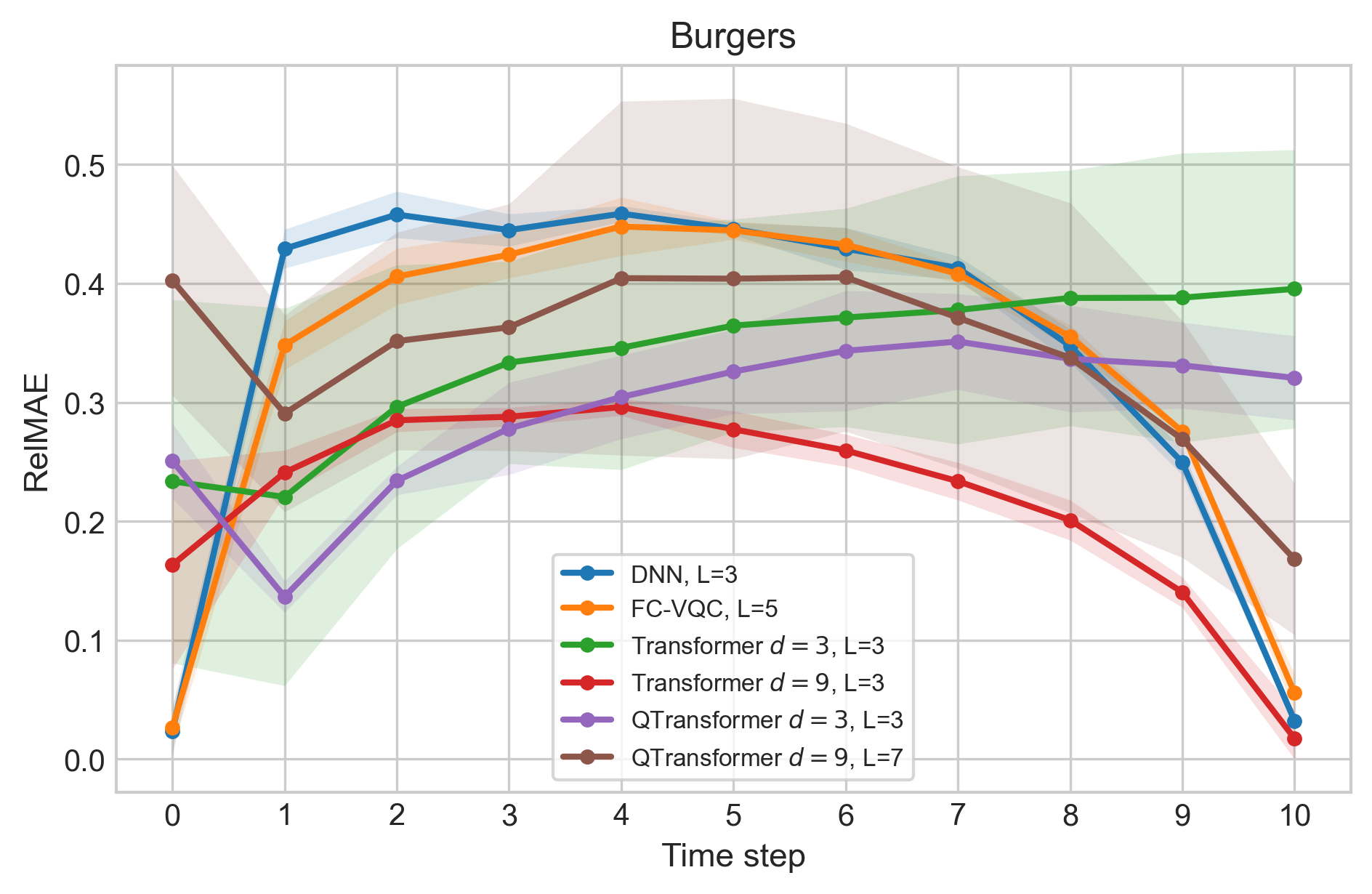}\\[-0.3em]
{\footnotesize (b) Burgers}
\end{minipage}
\hfill
\begin{minipage}[t]{0.31\textwidth}
\centering
\includegraphics[width=\linewidth]{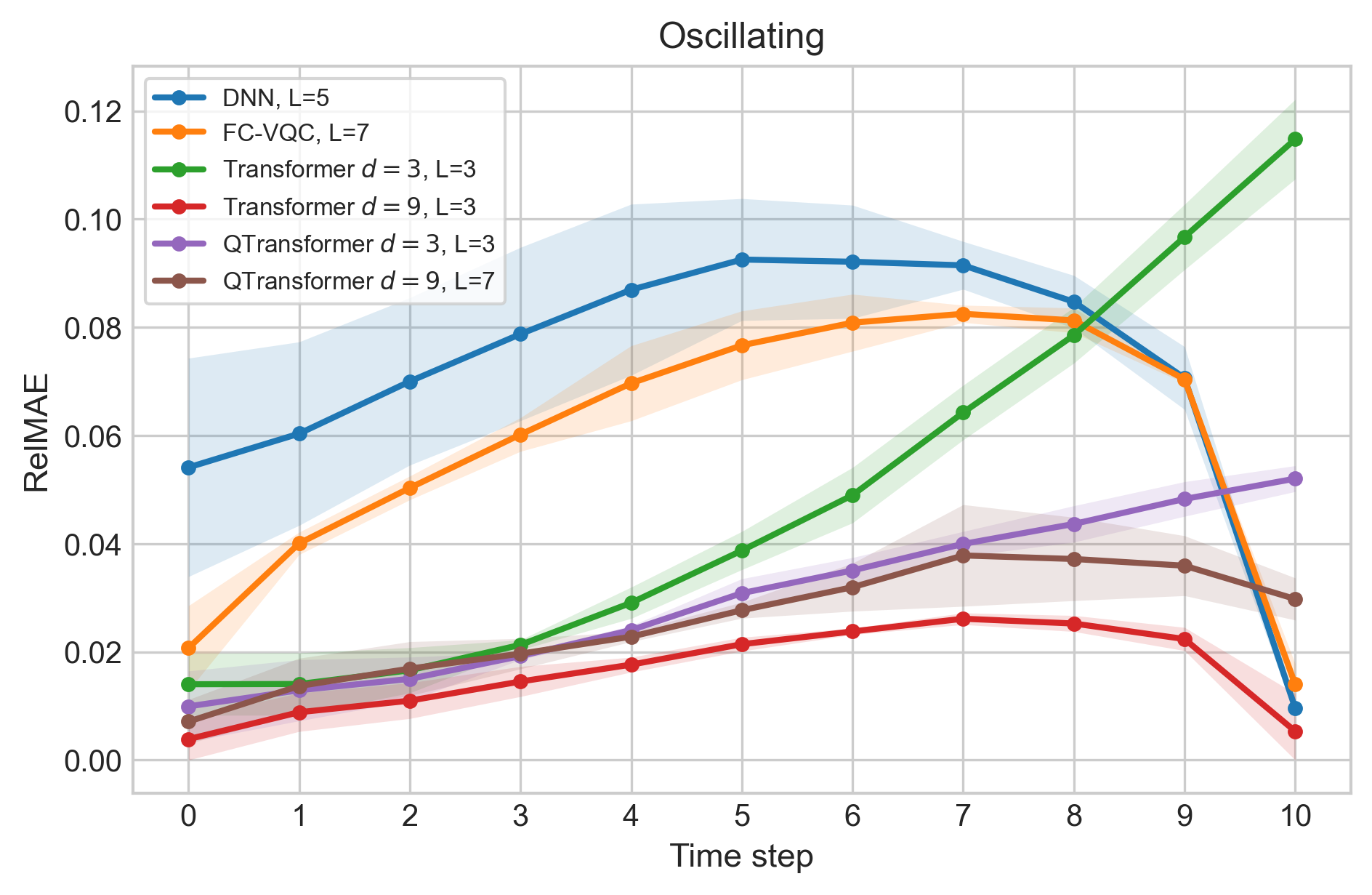}\\[-0.3em]
{\footnotesize (c) Oscillating}
\end{minipage}
\caption{Trajectory Relative MAE across time steps. Solid curves show
the mean over random seeds, and shaded regions indicate one standard
deviation.}
\vspace{-1em}
\label{fig:trajectory_relmae}
\end{figure*}

Table~\ref{tab:main_results} reports the mean trajectory RelMAE,
while Fig.~\ref{fig:trajectory_relmae} shows the corresponding
time-dependent errors with one standard deviation across random
seeds. The best-performing configurations exhibit the overall hierarchy:
Transformer (\(d_m=9\)) \(>\) QTransformer (\(d_m=9\))
\(>\) QTransformer (\(d_m=3\))
\(>\) Transformer (\(d_m=3\))
\(>\) FC-VQC \(>\) DNN, where \(>\) denotes lower RelMAE.
Overall, increasing the hidden dimension improves both Transformer
architectures, with the classical Transformer at \(d_m=9\) achieving
the lowest errors on most benchmarks.

QTransformer consistently improves over the non-attentive FC-VQC
baseline, demonstrating the benefit of causal self-attention for
learning BSDE gradient trajectories. At the compact hidden width
\(d_m=3\), QTransformer also consistently outperforms the classical
Transformer. To ensure a fair comparison, we note that at \(L=3\), 
QTransformer utilizes only 2,925 trainable parameters compared to 
the classical Transformer's 5,717. This confirms the quantum model's 
parameter-efficiency advantage is genuine and not an artifact of an 
under-parameterized classical baseline. Together with the stronger 
performance of the \(d_m=9\) models, these results indicate that both 
attention and hidden-width scaling contribute to improved accuracy.

\begin{table}[t]
\centering
\caption{Mean trajectory RelMAE on the three benchmark PDEs. Lower is
better. \(L\) denotes model layers. The hidden dimension
\(d_m=d_{\mathrm{model}}\) applies only to Transformer and
QTransformer.}
\label{tab:main_results}
\footnotesize
\setlength{\tabcolsep}{3.2pt}
\renewcommand{\arraystretch}{0.95}
\begin{tabular*}{\columnwidth}{@{\extracolsep{\fill}}llcccccc@{}}
\hline
Task & \(L\) & DNN & FC-VQC
& \multicolumn{2}{c}{Transformer}
& \multicolumn{2}{c}{QTransformer} \\
\cline{5-8}
& & & & \(d_m=3\) & \(d_m=9\) & \(d_m=3\) & \(d_m=9\) \\
\hline
BS
& 3 & \textbf{0.0246} & \textbf{0.0226} & \textbf{0.0148} & 0.0104 & 0.0148 & 0.0140 \\
& 5 & 0.0264 & 0.0235 & 0.0230 & \textbf{0.0099} & 0.0146 & 0.0136 \\
& 7 & 0.0265 & 0.0245 & 0.0271 & 0.0108 & \textbf{0.0142} & \textbf{0.0132} \\
\hline
Burg.
& 3 & \textbf{0.3402} & 0.3463 & \textbf{0.3392} & \textbf{0.2181} & \textbf{0.2933} & 0.2620 \\
& 5 & 0.3451 & \textbf{0.3309} & 0.4759 & 0.2200 & 0.2728 & 0.2617 \\
& 7 & 0.3576 & 0.3324 & 0.4747 & 0.2201 & 0.3067 & \textbf{0.2566} \\
\hline
Osc.
& 3 & 0.0876 & 0.0640 & \textbf{0.0486} & \textbf{0.0164} & \textbf{0.0300} & 0.0330 \\
& 5 & \textbf{0.0720} & 0.0610 & 0.0486 & 0.0486 & 0.0318 & 0.0265 \\
& 7 & 0.0910 & \textbf{0.0588} & 0.0486 & 0.0485 & 0.0272 & \textbf{0.0255} \\
\hline
\end{tabular*}
\end{table}

\subsection{Ablation Study}
\label{subsec:ablation}

\begin{table}[h]
\centering
\caption{QTransformer ablation comparing monolithic VQC and FC-VQC
trainable maps with \(d_m=d_{\mathrm{model}}=3\). Here \(L\) denotes
the number of Transformer layers. Entries are mean trajectory RelMAE;
lower is better.}
\vspace{-1em}
\label{tab:qtrans_ablation}
\footnotesize
\setlength{\tabcolsep}{5.5pt}
\renewcommand{\arraystretch}{1.05}
\begin{tabular*}{\columnwidth}{@{\extracolsep{\fill}}llcc@{}}
\hline
Task & \(L\) & Monolithic VQC & FC-VQC \\
     &       & \(d_m=3\)       & \(d_m=3\) \\
\hline
Black--Scholes
& 3 & 0.0145 & 0.0148 \\
& 5 & 0.0144 & 0.0146 \\
& 7 & 0.0153 & 0.0142 \\
\hline
Burgers
& 3 & 0.3914 & 0.2933 \\
& 5 & 0.3319 & 0.2728 \\
& 7 & 0.3432 & 0.3067 \\
\hline
Oscillating
& 3 & 0.0355 & 0.0300 \\
& 5 & 0.0365 & 0.0318 \\
& 7 & 0.0286 & 0.0272 \\
\hline
\end{tabular*}
\end{table}

Table~\ref{tab:qtrans_ablation} reports an ablation replacing each FC-VQC 
map inside QTransformer with a monolithic VQC map, keeping the structure 
and training protocol unchanged. The FC-VQC version consistently outperforms 
this monolithic baseline. Furthermore, for larger \(d_m\), the monolithic 
variant requires impractical \(d_m\)-qubit circuits per map. In contrast, 
FC-VQC scales by adding local blocks while keeping individual circuits 
small, validating both the accuracy and scalability of the design.

\section{Conclusion}
\label{sec:conclusion}

We proposed an FC-VQC-based Quantum Transformer BSDE solver for
high-dimensional parabolic PDEs. The method uses time--coordinate
tokens and causal self-attention to learn the adapted BSDE gradient
process. All trainable model parameters are contained within the
FC-VQC modules, while attention and structural operations remain
classical and parameter-free.

Across the benchmarks, the results show the overall hierarchy
Transformer (\(d_m=9\)) \(>\) QTransformer (\(d_m=9\))
\(>\) QTransformer (\(d_m=3\)) \(>\) Transformer (\(d_m=3\))
\(>\) FC-VQC \(>\) DNN, where \(>\) denotes lower RelMAE.
QTransformer consistently improves over FC-VQC and outperforms the
classical Transformer at \(d_m=3\), highlighting the benefit of causal
attention at compact hidden width. Increasing \(d_m\) improves both
Transformer architectures, with the classical Transformer at
\(d_m=9\) performing best overall.

The ablation study further shows that FC-VQC trainable maps can improve
QTransformer performance compared with monolithic VQC maps. Beyond accuracy, FC-VQC provides a practical scaling advantage: larger
\(d_m\) can be obtained by adding local VQC blocks, whereas the
monolithic variant requires \(d_m\)-qubit circuits that may become
impractical for larger hidden dimensions.

These initial results rely on ideal simulations and three random seeds. Future work will incorporate statistical significance testing across more 
seeds, evaluate finite-shot measurement and hardware noise impacts, and 
empirically measure gradient variances to substantiate the qualitative 
barren-plateau mitigation claim. We will also extend models to larger 
hidden dimensions under strictly matched-parameter budgets to further 
isolate genuine quantum inductive biases.

\bibliographystyle{IEEEtran}
\bibliography{refs}

\end{document}